\documentclass[aps,twocolumn,showpacs,superscriptaddress]{revtex4-1}
\usepackage{amsfonts}
\usepackage{amsmath}
\usepackage{amssymb}
\usepackage{lipsum}
\usepackage{graphicx}
\usepackage{bm}
\usepackage{natbib}
\usepackage{float}
\usepackage{multirow}
\usepackage{ulem}
\usepackage[dvipsnames]{xcolor}
\newcommand{\sgn}{\rm sign}

\begin{document}
\preprint{APS/123-QED}
\title{Magnetic, charge, and transport properties of graphene nanoflakes}
\author{V. S. Protsenko}
\affiliation{M. N. Mikheev Institute of Metal Physics of Ural Branch of Russian Academy of Sciences, S. Kovalevskaya Street 18, 620990 Yekaterinburg, Russia}
\affiliation{Theoretical Physics and Applied Mathematics Department, Ural Federal University, 620002 Yekaterinburg, Russia}
\author{A. A. Katanin}
\affiliation{Center for Photonics and 2D Materials, Moscow Institute of Physics and Technology, Institutsky lane 9, Dolgoprudny, 141700, Moscow region, Russia}
\affiliation{M. N. Mikheev Institute of Metal Physics of Ural Branch of Russian Academy of Sciences, S. Kovalevskaya Street 18, 620990 Yekaterinburg, Russia}
\date{\today}
\begin{abstract}
We investigate magnetic, charge and transport properties of hexagonal graphene nanoflakes (GNFs) connected to two metallic leads by using the functional renormalization group (fRG) method. 
The interplay between the on-site and long-range interactions leads to a competition of semimetal (SM), spin density wave (SDW), and charge-density-wave (CDW) phases. The ground-state phase diagrams are presented for the GNF systems with screened realistic long-range electron interaction [T. O. Wehling, et. al., Phys. Rev. Lett. {\bf 106}, 236805 (2011)], as well as uniformly screened long-range Coulomb potential $\propto 1/r$. We demonstrate that the realistic screening of  Coulomb interaction by $\sigma$ bands causes moderate (strong) enhancement of critical long-range interaction strength, needed for the SDW (CDW) instability, compared to the results for the uniformly screened Coulomb potential. This enhancement gives rise to a wide region of stability of the SM phase for realistic interaction,
such that freely suspended GNFs are far from both SM-SDW and SM-CDW phase-transition boundaries and correspond to the SM phase. 
Close relation between the linear conductance and the magnetic or charge states of the systems is discussed. A comparison of the results with those of other studies on GNFs systems and infinite graphene sheet is presented.
\end{abstract}
\maketitle
\section{Introduction}
Recent progress in the fabrication of nanostructures has allowed reducing the dimension of graphene plane from two to a zero-dimensional system -- graphene nanoflake (GNF)~\cite{Luo_2011,Guttinger_2012,Snook_2011}. In addition to being a promising building block for nanoelectronics and spintronic devices~\cite{Yazyev_2010,Ezawa_2009}, GNFs are interesting in many aspects of a fundamental point of view. On one hand, the electronic structure of these finite graphene nanostructures can be qualitatively different from that of graphene~\cite{Yazyev_2010,Rozhkov_2011}. In particular, due to 
appearance of edge states, which do not arise in an infinite graphene sheets, the electronic properties of GNF may depend significantly on the edge geometry~\cite{Ritter_2009,Yazyev_2010,Rozhkov_2011,Fernandez_2007}. On the other hand, even small GNF systems with less than one hundred atoms demonstrate the occurrence of charge and magnetic instabilities, which are analogous to those of single-layer graphene~\cite{DCA,Valli_2016,Valli_2018,Valli_2019}.

In an infinite graphene sheet, 
both short- and long-range electron-electron interactions play an important role, favoring 
spin density wave  (SDW) \cite{UInf1,UInf2,UInf3,UInf4,Tang_2018,Ulybyshev_2013,Buividovich_2018} or charge density wave (CDW) 
\cite{Strouthos,Khveschenko,Gusynin,Murthy,Drut,Khveschenko_d,Gusynin_d,Gonzalez,Katanin,Buividovich_2018} correlations. One may expect the same types of correlations to be relevant for GNF. In particular, GNF clusters up to 96 sites with on-site $U$ and nearest-neighbor $V$ interactions were investigated 
in Ref.~\cite{DCA} within the dynamical cluster approximation (DCA). It was found that the competition between short- and long-range interactions gives rise to a nontrivial phase diagram, which includes the transitions between semimetal (SM), SDW, and CDW phases. An increase of $U$ (for $V=0$) induces a phase transition from the SM to SDW state, while sufficiently large nearest-neighbor interaction $V$ leads to the CDW ground state. At and close to half-filling the transition to the SDW state was also investigated within the dynamic mean-field theory (DMFT) approach for GNF system with 54 atoms coupled to leads~\cite{Valli_2016,Valli_2018,Valli_2019}. 
The emergence of magnetism in GNF clusters, induced by electron-electron interaction, including formation of finite magnetic moments, was also predicted within the mean-field~\cite{Fernandez_2007, Fujita_1996, Yazyev_2010, Szalowski_2015} and density functional theory (DFT) calculations~\cite{Fernandez_2007,Kabir_2014,Ganguly_2017}. The  
edge magnetization was thoroughly studied, especially for GNF with zigzag edges~\cite{Yazyev_2010,Rozhkov_2011,Kabir_2014,Hagymasi_2018, Luo_2014,Shi_2017,Raczkowski_2017}. 

To date, the vast majority of studies on magnetic properties of graphene nanosystems  
considered only local and nearest-neighbor electron-electron interactions~\cite{Yamashiro_2003,Chacko_2014,Zhu_2006, Chacko_2014}. However, it is well-known that the long-range part of the electron-electron Coulomb interaction is not screened in single-layer graphene. This suggests that the interactions beyond nearest neighbor distance can also have a crucial impact on physics of graphene nanosystems, which is confirmed by ab initio calculations~\cite{Hadipour_2018}. Although there are several papers focused on investigating the effects of long-range interaction in graphene nanoflakes beyond the nearest-neighbor interactions {\cite{DCA,Ozdemir_2016,Wunsch_2008,Luo_2014,Sheng_2013}}, most of them are limited to the $1/r$~\cite{Wunsch_2008,Sheng_2013} or $1/\sqrt{r^{2}+a^{2}}$~\cite{Luo_2014} dependence of non-local potential on distance $r$, the parameter $a$ accounts for finite radius of graphene $\pi$ orbitals. 

At the same time, realistic non-local interaction in graphene 
have been determined by accurate first-principles calculations~\cite{Wehling_2011}. 
At intermediate distances, this realistic potential differs significantly from the standard Coulomb potential~\cite{Wehling_2011,
Ulybyshev_2013} due to screening 
of interaction by $\sigma$ orbitals. It has been predicted that this difference of the potentials results in the shift of the critical value of the semimetal-insulator phase transition in comparison to the previous estimates \cite{Ulybyshev_2013}. Corresponding position of SDW instability was obtained by the hybrid quantum Monte Carlo (QMC) simulations~\cite{Buividovich_2019,Ulybyshev_2013,Smith_2014} and was shown to correspond to the dielectric permittivity $\epsilon<1$. These considerations show that the magnetic properties of GNF with the standard bare $1/r$ form of the Coulomb potential may differ significantly from the ones for the realistic model. 
Thus, it is highly desirable to analyze the competition between different magnetic and charge instabilities in GNFs with the realistic model of non-local interactions. 

Another important issue is the impact of electron-electron interaction effects on electron transport of GNF systems. Most studies on this issue are based on either the (dynamic) mean-field or DFT approaches, which do not fully include electron correlation effects. Furthermore, despite the attention paid to the effects inherent to mesoscopic systems in general (e.g., Coulomb blockade~\cite{Ezawa_2008, Weymann_2012} and quantum interference~\cite{Sahin_2008,Valli_2018,Valli_2019}), little attention has been focused on the relationship between the electron transport and charge (or spin) correlations in GNF systems (see e.g.~\cite{Weymann_2012,Luo_2014}). Understanding this relationship may be important for application of the GNF systems in the development of spin filters and other spintronic devices.

In the present paper, we study the magnetic, charge, and transport properties of hexagonal GNFs of different sizes with realistic non-local interaction. The effects of both short- and long-range electron-electron interactions are studied at zero temperature by using the functional renormalization group method. Considering screening of the on-site and non-local components of interaction independently we obtain the phase diagrams of the GNFs. We obtain the SDW (CDW) phases 
at sufficiently strong local (non-local) interactions as has been discussed previously. We show that for uniformly screened Coulomb interaction the position of the obtained instabilities is in agreement with previous studies, but for realistic non-local interaction their position in the phase diagram is somewhat changed.
In particular, in the latter case phase diagrams show the presence of a wide region with no instability.
Although the position of spin instability for realistic non-local interaction was studied previously in Ref. \cite{Ulybyshev_2013} for sufficiently large graphene sheet, this instability in graphene nanoflakes, as well as the parameters of charge instability for graphene nanoflakes and infinite graphene sheet in the presence of both, local and non-local interaction, to our knowledge were not determined previously. As we argue in the present paper, the CDW instability is much stronger affected by the screening of Coulomb interaction by $\sigma$ bands, than the SDW instability. 
We also present results for the linear conductance of the GNF systems and clarify features of the conductance associated with the
transitions between different magnetic regimes.

The paper is organized as follows. In Sect. II, after presenting the model Hamiltonian for GNF systems we describe the functional renormalization group method. In Sect. III we consider the results for a purely local interaction, present phase diagrams with account of non-local interaction, discuss stability of the CDW and SDW order, and present results for the linear conductance. 
Finally, Sect. IV  summarizes our main results and  {presents} conclusions.
\section{Model and Method}
\begin{figure}[b]
		\center{\includegraphics[width=0.8\linewidth]{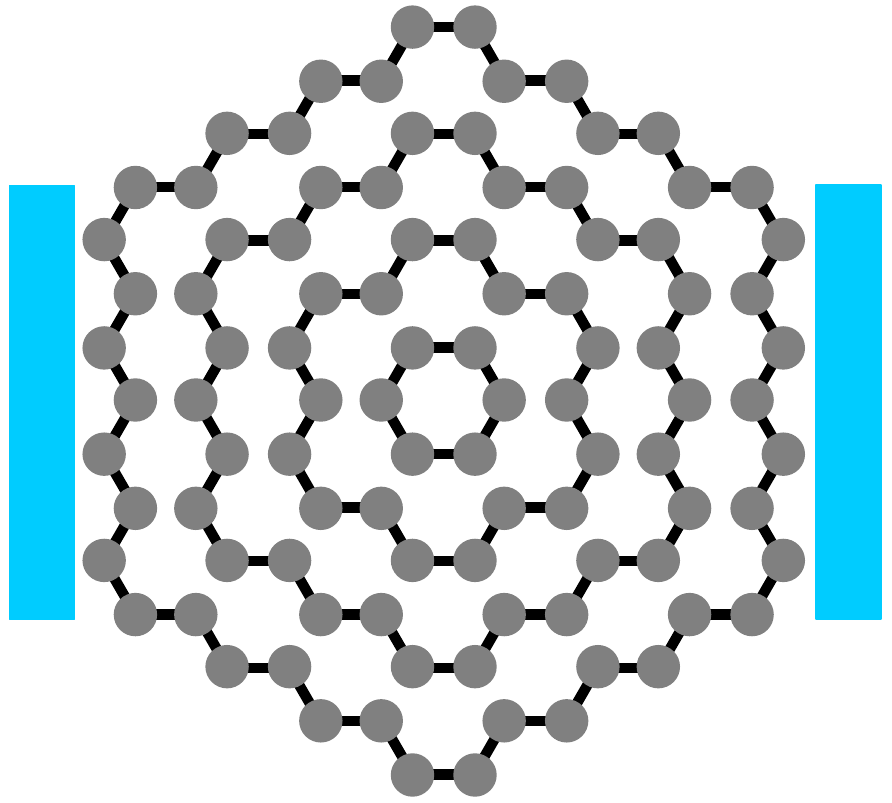}}
		\caption{{(Color online)} GNF-$N$ systems considered in the paper (regions bounded by closed concentric lines, corresponding to $N=6,24,54,96$ from inner to outer line). The left and right leads are shown schematically by rectangles.}
\label{SystemsPic}
\end{figure}

We consider the systems, consisting of a graphene nanoflake (GNF-$N$) with $N$ atoms connected to two metallic leads (see Fig.~\ref{SystemsPic} for zigzag edge geometry). The total Hamiltonian of the GNF-$N$ system can be written as
\begin{equation}
	\mathcal{H}= \mathcal{H}_{\rm GNF}+\mathcal{H}_{\rm leads}+\mathcal{H}_{\rm T}.
	\label{TotalHamiltonian}
\end{equation}
The first term describes the isolated graphene nanoflake,
\begin{multline}
 \mathcal{H}_{\rm GNF}=\sum_{\sigma}\sum_{i\in A}\epsilon_{\sigma}^{A}n_{i,\sigma}+\sum_{\sigma}\sum_{i\in B}\epsilon_{\sigma}^{B}n_{i,\sigma}\\-t\sum_{<ij>,\sigma}d^{\dagger}_{i,\sigma}d_{j,\sigma}+\dfrac{1}{2}\sum_{i,j}U_{ij}\left(n_{i}-1\right)\left(n_{j}-1\right).
 \label{GNFHamiltonian}
\end{multline}
Here, $d^{\dagger}_{i,\sigma}$ $\left(d_{i,\sigma}\right)$ is a creation (annihilation) operator of an electron at the lattice site $i$ {of $A$ or $B$ sublattice} with a spin index $\sigma=\pm 1/2$ (or $\sigma=\uparrow,\downarrow$), $n_{j,\sigma}=d^{\dagger}_{j,\sigma}d_{j,\sigma}$ and $n_{j}=n_{j,\uparrow}+n_{j,\downarrow}$. The on-site energy parameters  are chosen to be $\epsilon_{\sigma}^{A(B)}=\pm\left(\delta-h\sigma\right)$, 
the parameter $\delta$ and {magnetic field} $h$ are introduced in order to explicitly break the spin and sublattice symmetry of the GNF-$N$, $t=2.7$~eV 
is the nearest-neighbor hopping parameter and summation in the third term of Eq.~(\ref{GNFHamiltonian}) is taken over nearest neighbor sites. 
 The last term in Eq.~(\ref{GNFHamiltonian}) describes the electron-electron interactions with the potential $U_{ij}$  
 that includes both on-site $U=U_{ii}$ and non-local $U_{i\ne j}$ contributions. 

In the following we mainly use the form 
$U_{i\ne j}=U^{*}_{ij}/\epsilon_{\rm nl}$ of the non-local interaction, 
where $U^{*}_{ij}$ is the realistic non-local potential of Ref.~\cite{Wehling_2011}, which accounts for the screening of Coulomb interaction by $\sigma$ orbitals. At distances larger than the distance between third-nearest-neighbor lattice sites $r_{ij}>r_{03}=2a$ ($a=0.142$ nm is graphene's lattice constant) the realistic potential is approximated by $U^*_{ij}=1/(\epsilon_{\rm eff} r_{ij})$ with effective dielectric permittivity {$\epsilon_{\rm eff}=1/(U^{*}_{03}r_{03})\approx 1.41$}{, as in Ref.~\cite{Ulybyshev_2013}}. In this case, independent variation of the parameters $U$ 
and $\epsilon_{\rm nl}$ allows us to study the interplay between the on-site and non-local parts of the interactions. 
According to Ref.~\cite{Wehling_2011}, 
$U=U_r=9.3~{\rm eV}{\approx}3.44t$ corresponds to the realistic on-site interaction in graphene. 

The second part  $\mathcal{H}_{\rm leads}$ of the Hamiltonian~(\ref{TotalHamiltonian}) describes the two equivalent metallic leads,
\begin{equation}
\mathcal{H}_{\rm leads}=\sum_{k,\alpha,\sigma}\epsilon_{k}^{\phantom{\dagger}}c^{\dagger}_{k,\alpha,\sigma}c^{\phantom{\dagger}}_{k,\alpha,\sigma}.
\label{H_leads}
\end{equation}
Here, $c^{\dagger}_{k,\alpha,\sigma}$ $\left(c_{k,\alpha,\sigma}\right)$ is a creation (annihilation) operator of an electron with the state $k$ and spin $\sigma$ in the left $(\alpha=L)$ or right $(\alpha=R)$ lead, $\epsilon_{k}$ represents the single-particle energy. \par
The last term in Eq.~(\ref{TotalHamiltonian}) describes connection between the GNF-$N$ and leads, following Ref. \cite{Valli_2019} we consider it in the form
\begin{equation}
\mathcal{H}_{\rm T}=-\sum_{\sigma,k,\alpha,i_\alpha}\left(V^{\phantom{\dagger}}_{i_\alpha, k, \alpha} c^{\dagger}_{k,\alpha,\sigma}d^{\phantom{\dagger}}_{i_\alpha,\sigma}+\text{H.c.}\right),
\label{H_T}
\end{equation}
where $V_{i,k,\alpha}$ is the coupling matrix element between the $i$-th site of GFN-$N$ and the $k$-th state of the lead $\alpha$, and summation is performed over sites $i_\alpha$ that are closest to the lead $\alpha$. 
\par

To reveal the formation of SDW and CDW phases of GNF-$N$ we calculate the average relative staggered magnetization 
\begin{equation}
S_{\rm st} =\left(\langle N_{A,\uparrow} \rangle+\langle N_{B,\downarrow}\rangle-\langle N_{A,\downarrow} \rangle -\langle N_{B,\uparrow} \rangle\right)/N
\end{equation}
and the average relative difference between the occupation of A and B sublattices
\begin{equation}
\Delta_{\rm st}=\left(\langle N_{B,\uparrow} \rangle+\langle N_{B,\downarrow}\rangle-\langle N_{A,\uparrow} \rangle -\langle N_{A,\downarrow} \rangle\right)/N,
\end{equation}
{which have maximal value of $1$, ${\langle}N_{A(B),\sigma}{\rangle}=\sum_{i\in A(B)} {\langle}n_{i,\sigma}{\rangle}$. {It is worth noting that 
due to the particle-hole symmetry of the Hamiltonian~(\ref{GNFHamiltonian}) the total average occupation of the GNF-$N$ is automatically fixed to half-filling, $ \sum_{\sigma} \left(\langle N_{A,\sigma} \rangle+\langle N_{B,\sigma} \rangle\right)=N$ even in presence of electron-electron interactions.}}
The  average occupation of a lattice site $i$ for spin $\sigma$ can be calculated at temperature $T=0$ as
\begin{equation}
\langle n_{j,\sigma}\rangle=
\int{\dfrac{d\omega}{2\pi} e^{i\omega 0^{+}}
	\mathcal{G}
	_{jj,\sigma}}(i \omega).
\label{occ}
\end{equation}
Here $\mathcal{G}(i \omega)$ is the Green's function corresponding to the Hamiltonian~(\ref{TotalHamiltonian}) projected onto the states of GNF-$N$. 

{To determine the Green's function $\mathcal{G}(i \omega)$ we use the functional renormalization group (fRG) technique \cite{Salmhofer_1,Metzner,Karrasch_2006}. This technique introduces cutoff parameter $\Lambda$, specified below in Eq. (\ref{Hyb}), such that the physical Green's function is obtained} in the end of the fRG flow, i.e. for the corresponding cutoff parameter $\Lambda=0$, $\mathcal{G}(i \omega)=\mathcal{G}^{\Lambda=0}(i \omega)$. By using the Dyson equation and the projection technique the cutoff-dependent Green's function  $\mathcal{G}^\Lambda(i \omega)$ can be written as
\begin{equation}
\mathcal{G}^{\Lambda}(i \omega)=\left[(\mathcal{G}^{\Lambda}_{0}(i \omega))^{-1}-\Sigma_{\rm leads}-\Sigma^{\Lambda}\right]^{-1},
\end{equation}
where $\mathcal{G}^{\Lambda}_{0}(i \omega)$ is the cutoff dependent bare Green's function, $\Sigma_{\rm leads}$ describes the coupling between the GNF-$N$ and leads, and $\Sigma^{{\Lambda}}$ is the self-energy of the interacting $(U_{ij}\ne0)$ system. Following Ref.~\cite{Valli_2019}, for $\Sigma_{\rm leads}$ we take into account only {diagonal (with respect to site indices)} hybridization processes and use the wide band limit approximation~\cite{Jauho_1994}. This leads to $\Sigma_{\rm leads}=-i\Gamma\sgn(\omega)$ for each site of the GNF-$N$ connected to leads, where $\Gamma {\propto |V|^2} \rho_{\text{lead}}$ is an energy independent hybridization strength {and $\rho_{\text{lead}}$ is the density of states in the leads.} \par
The self-energy $\Sigma^{\Lambda}$ 
can be obtained from an infinite hierarchy of differential flow equations
for the cutoff-parameter $\Lambda$ dependent self-energy $\Sigma^\Lambda$ and the $n$--particle vertices $\Gamma_{2n}^{\Lambda}$, $n\geq2$ \cite{Salmhofer_1,Metzner}.  Truncating the fRG flow equations by neglecting the flow of the vertex functions with $n\geq3$  and discarding the frequency dependence of vertices leads to a closed system of the flow equations for the $\Sigma^{\Lambda}$ and the two-particle vertex $\Gamma_{4}^{\Lambda}$~\cite{Metzner,Karrasch_2006}.

In the present study, we use the coupled-ladder approximation~\cite{Bauer_2014,Weidinger_2017} to this closed set of fRG equations, which makes numerical calculations feasible for the systems under consideration. This approximation consists in decomposition of the two-particle vertex $\Gamma_{4}^{\Lambda}$ into the particle-particle ($P^{\Lambda}$), the exchange particle-hole ($X^{\Lambda}$), and the direct particle-hole ($D^{\Lambda}$) channels
\begin{align}
&\Gamma_{4}^{\Lambda}\left(j_{1}^{'},j_{2}^{'},j_{1}^{\phantom{'}},j_{2}^{\phantom{'}},\bm{\sigma}\right)=I\left(j_{1}^{'},j_{2}^{'},j_{1}^{\phantom{'}},j_{2}^{\phantom{'}},\bm{\sigma}\right)\notag\\&+P^{\Lambda}\left(j_{1}^{'},j_{1}^{\phantom{'}},\bm{\sigma}\right)\delta_{j_{1}^{'}j_{2}^{'}}\delta_{j_{1}^{\phantom{'}}j_{2}^{\phantom{'}}}+X^{\Lambda}\left(j_{1},j_{2},\bm{\sigma}\right)\delta_{j_{1}^{'}j_{1}^{\phantom{'}}}\delta_{j_{2}^{'}j_{2}^{\phantom{'}}}\notag\\&+D^{\Lambda}\left(j_{2},j_{1},\bm{\sigma}\right)\delta_{j_{2}^{'}j_{1}^{\phantom{'}}}\delta_{j_{1}^{'}j_{2}^{\phantom{'}}},
\label{efRG}
\end{align}
and splitting the flow equation for the two-particle vertex $\Gamma_{4}^{\Lambda}$ into the equations for these individual channels. Term $I$ in Eq.~(\ref{efRG}) is the antisymmetrized bare interaction and $\bm{\sigma}=(\sigma_{1}^{'},\sigma_{2}^{'},\sigma_{1},\sigma_{2})$ is a multi-index. The fRG equations for the self-energy $\Sigma^{\Lambda}$ and for the vertices $P^{\Lambda}$, $X^{\Lambda}$, $D^{\Lambda}$ can be written in the form
\begin{align}
\partial_{\Lambda}\Sigma^{\Lambda}&=-\int{\dfrac{d\omega}{2\pi} e^{i\omega 0^{+}}\mathcal{S}^{\Lambda}\left(i\omega\right)\circ\Gamma_{4}^{\Lambda}},\label{SEeq}\\
\partial_{\Lambda}P^{\Lambda}&=\int{\dfrac{d\omega}{2\pi}\Gamma_{p}^{\Lambda}\circ\mathcal{S}^{\Lambda}\left(i\omega\right)\circ\mathcal{G}^{\Lambda}\left(-i\omega\right)\circ\Gamma_{p}^{\Lambda}},\label{Peq}\\
\partial_{\Lambda}X^{\Lambda}&=-\int{\dfrac{d\omega}{2\pi}\Gamma_{x}^{\Lambda}\circ\left(\mathcal{S}^{\Lambda}\left(i\omega\right)\circ\mathcal{G}^{\Lambda}\left(i\omega\right)\right.}\notag\\&\left.+\mathcal{G}^{\Lambda}\left(i\omega\right)\circ\mathcal{S}^{\Lambda}\left(i\omega\right)\right)\circ\Gamma_{x}^{\Lambda},\label{Xeq}
\end{align}
\begin{align}
    \partial_{\Lambda}D^{\Lambda}&=\int{\dfrac{d\omega}{2\pi}\Gamma_{d}^{\Lambda}\circ\left(\mathcal{S}^{\Lambda}\left(i\omega\right)\circ\mathcal{G}^{\Lambda}\left(i\omega\right)\right.}\notag\\&\left.+\mathcal{G}^{\Lambda}\left(i\omega\right)\circ\mathcal{S}^{\Lambda}\left(i\omega\right)\right)\circ\Gamma_{d}^{\Lambda},\label{Deq}
\end{align}
where
\begin{align}
\Gamma_{p}^{\Lambda}\left(j_1,j_2,\bm{\sigma}\right)&=P^{\Lambda}\left(j_1,j_2,\bm{\sigma}\right)+I\left(j_1,j_1,j_2,j_2,\bm{\sigma}\right)\\
\Gamma_{x}^{\Lambda}\left(j_1,j_2,\bm{\sigma}\right)&=X^{\Lambda}\left(j_1,j_2,\bm{\sigma}\right)+I\left(j_1,j_2,j_1,j_2,\bm{\sigma}\right)\\
\Gamma_{d}^{\Lambda}\left(j_1,j_2,\bm{\sigma}\right)&=D^{\Lambda}\left(j_1,j_2,\bm{\sigma}\right)+I\left(j_1,j_2,j_2,j_1,\bm{\sigma}\right)
\end{align}
for $j_1\ne j_2$ and
\begin{multline}
\Gamma_{f}^{\Lambda}\left(j_1,j_1,\bm{\sigma}\right)=P^{\Lambda}\left(j_1,j_1,\bm{\sigma}\right)+X^{\Lambda}\left(j_1,j_1,\bm{\sigma}\right)+\\ D^{\Lambda}\left(j_1,j_1,\bm{\sigma}\right)+I\left(j_1,j_1,j_1,j_1,\bm{\sigma}\right)
\end{multline}
for $j_1=j_2$, $f=p,x,d$. In Eqs.~(\ref{SEeq})-(\ref{Deq}) $\mathcal{S}^{\Lambda}$ is the single-scale propagator
\begin{equation}
\mathcal{S}^{\Lambda}=\mathcal{G}^{\Lambda}\partial_{\Lambda}\left(\mathcal{G}^{\Lambda}_{0}\right)^{-1}\mathcal{G}^{\Lambda}.
\end{equation}
and "$\circ$" denotes summations over intermediate site and spin indexes, which  perform according to the standard diagrammatic rules. 
 
To introduce the cutoff-parameter $\Lambda$ we use the reservoir cutoff scheme~\cite{Karrasch_2010} in the form
\begin{equation}
\mathcal{G}^{\Lambda}_{0}(i \omega)=\left(\mathcal{G}^{-1}_{0}(i \omega)+iI_{N}\Lambda\sgn(\omega) \right)^{-1},
\label{Hyb}
\end{equation}
where $I_{N}$ is the $N\times N$ identity matrix, $\mathcal{G}_0(i\omega)$ is the bare Green's function, corresponding to the single-particle part of the Hamiltonian~(\ref{GNFHamiltonian}) (i.e $\mathcal{H}_{\rm GNF}$ with $U_{ij}=0$). The value of $\Sigma^{\Lambda}$ obtained at the end of the fRG flow (for $\Lambda\rightarrow 0$) corresponds to the physical self-energy $\Sigma$ of the interacting system. \par
The linear conductance $G=e^{2}/h \sum_{\sigma}{\mathcal T}_{\sigma}(\omega\rightarrow 0)$ at $T=0$, where the transmission function ${\mathcal T}_{\sigma}(\omega)$ in our case can be written in the form~\cite{Valli_2018,Valli_2019}
\begin{equation}
{\mathcal T}_{{\sigma}}(\omega)=4\Gamma^{2}\sum_{i,j}\left|\mathcal{G}^{r}_{ij,\sigma}(\omega)\right|^{2}.
\label{G(Vg)}
\end{equation}
Here $\mathcal{G}^{r}(\omega)=\mathcal{G}^{\Lambda=0}(i\omega\rightarrow\omega+i 0^{+})$ is the retarded Green's function of the GNF-$N$ and summation over site indexes $i$ ($j$) is restricted to sites of GNF-$N$ connected to the left (right) lead. Note that Eq.~(\ref{G(Vg)})  assumes that only local hybridization processes affect on the transmission~\cite{Valli_2019}.
\section{Results}


In the following calculations we set $\delta=0.0185t$, $\Gamma=0.02t$, $T=0$, and, unless otherwise stated, $h=\delta$.
\subsection{Zigzag-edge GNF-$N$ with purely local interaction
}
\begin{figure}[t]
	\center{\includegraphics[width=1\linewidth]{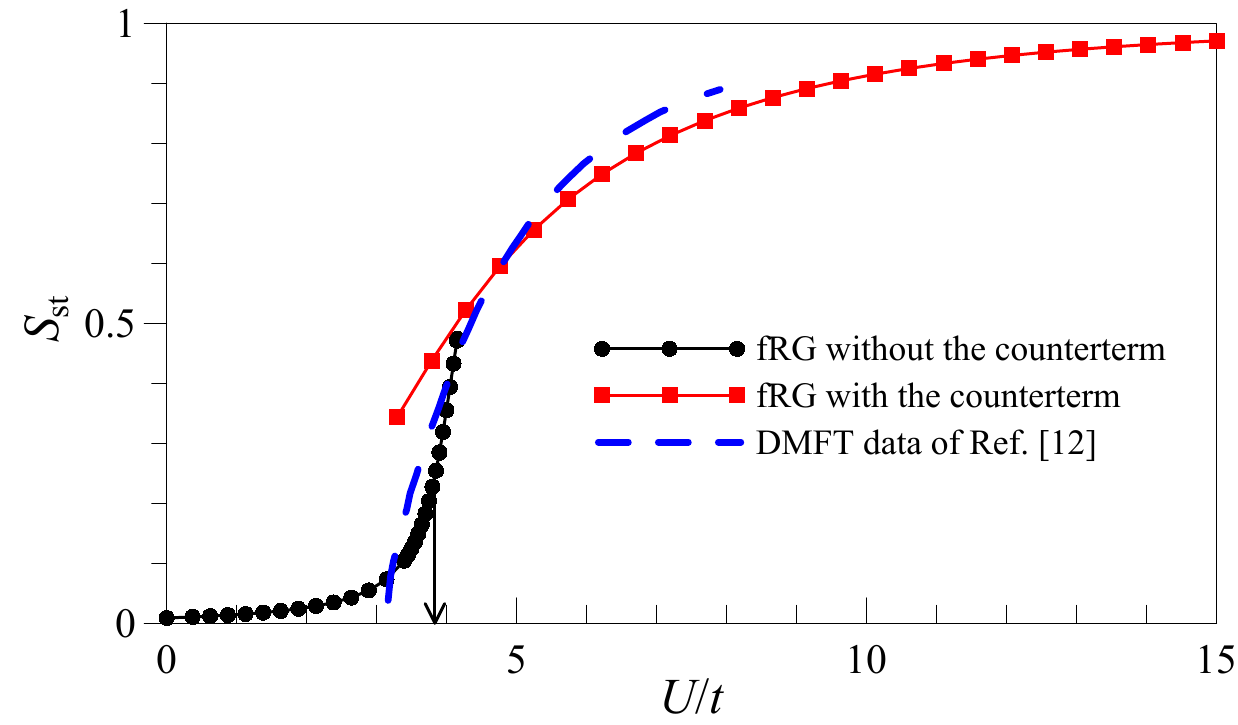}}
		\caption{(Color online) The average {relative} staggered magnetization $S_{\rm st}$ as a function of $U/t$ for the GNF-54 system with $U_{i\ne j}=0$. The square (circle) symbols correspond to the fRG results with (without) the counterterm. The dashed line represents the DMFT data of Ref.~\cite{Valli_2018}. The arrow indicates $U=U_{c}^{54}\approx 3.83t$ corresponding to $S_{\rm st}=1/4$.}
	\label{GNF54Local}
\end{figure}

Let us first consider the zigzag-edge GNF-$N$ systems with a purely local (on-site) interaction $U$, when $U_{i\ne j}=0$. In Fig.~\ref{GNF54Local}, the fRG results for the average relative staggered magnetization $S_{\rm st}$ are shown for the GNF-54 system. In the limit $U\ll t$, the magnetization of the GNF-$N$ system is small but nonzero due to the presence of the finite magnetic field. 
With increasing $U/t$, the magnetization increases monotonously, indicating the formation of the SDW order. For sufficiently large $U/t$, the vertices obtained from fRG equations diverge. 

Similarly to Ref.~\cite{IILM} we have found that the convergence of the vertices obtained from fRG equations can be achieved by applying the counterterm technique. 
The counterterm (which corresponds in our case to introducing auxiliary magnetic field $\tilde{h}=1.5t$, switched off linearly with $\Lambda$ starting from the scale $\Lambda_{c}=0.1t$) allows us continuing the $S_{\rm st}(U/t)$ dependence beyond the point at which the fRG approach without the counterterm breaks down, see Fig. 2. However, for smaller $U/t$ the fRG results with and without the conterterm are different from each other. Apparently, this is due to the unphysical spin-splitting of the self-energy in the fRG approach with a counterterm, which does not allow us to correctly reproduce the SM state of the system. In Fig.~\ref{GNF54Local} we also compare fRG results to the DMFT results of Ref.~\cite{Valli_2018} in the absence of magnetic field. One can see that for substantial (small) $S_{\rm st}$
the fRG approach with (without) the counterterm provides a reasonable agreement with the DMFT data of Ref.~\cite{Valli_2018}.

Since we are interested in the position of phase transitions to SDW (CDW) phases, rather than study of the regions deeply inside these phases, in the following we restrict ourselves to fRG approach without the counterterm.

\begin{figure}[t]
	\center{\includegraphics[width=0.95\linewidth]{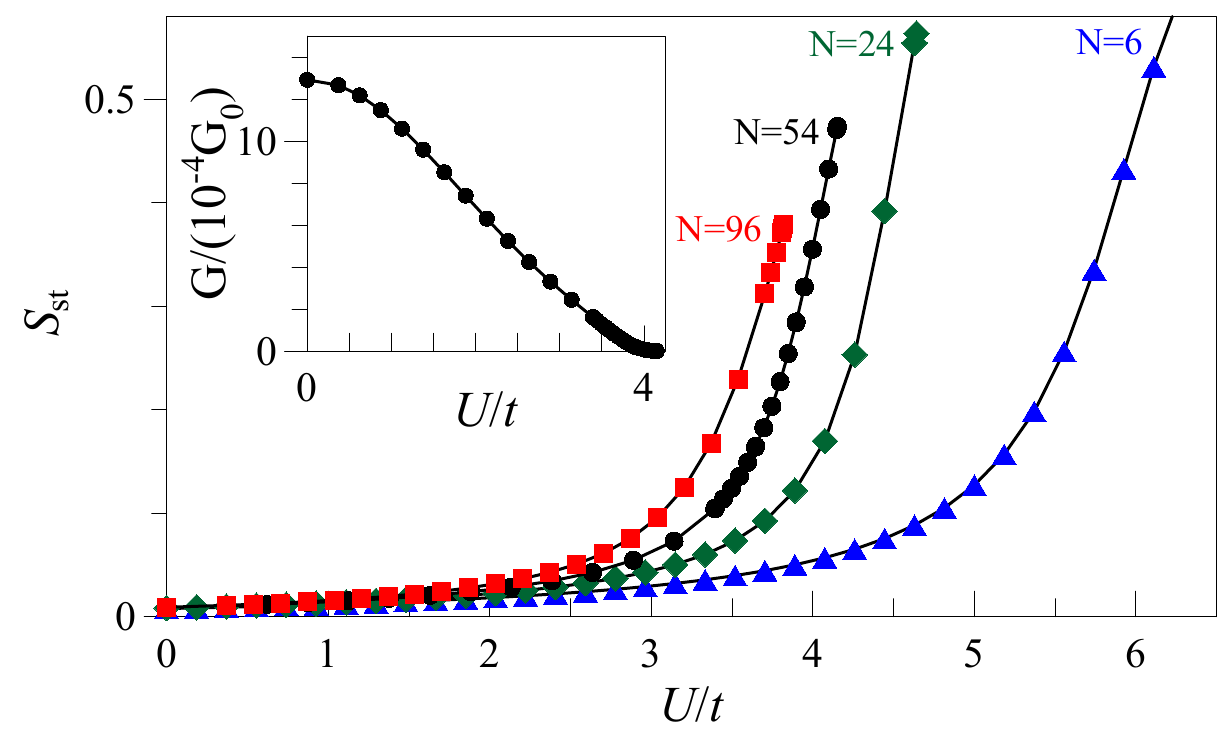}}
	\caption{(Color online) The average {relative} staggered magnetization $S_{\rm st}$ as a function of $U/t$ for the GNF-$N$ systems with $U_{i\ne j}=0$ and $N=6$ (blue triangles), $N=24$ (green diamonds), $N=54$ (black circles), $N=96$ (red squares). Inset: The linear conductance $G$ as a function of $U/t$ for the GNF-54 system with $U_{i\ne j}=0$, {$G_0=e^{2}/h$} {is the conductance quantum {per spin projection}}. 
	}
	\label{GNFnLocal}
\end{figure}

Figure~\ref{GNFnLocal} shows the average {relative} staggered magnetization $S_{\rm st}$ as a function of $U/t$ for a series of the GNF-$N$ systems of different sizes. In our case of a finite {small} magnetic field, {which is introduced to break explicitly the spin symmetry}, the transition from the SM ($S_{\rm st}{\rightarrow 0}$) to SDW ($S_{\rm st}\approx 1$) phase is smoothed.
To {find the position of phase} transition between the SM and SDW phases we define a characteristic local interaction $U^{N}_{c}$ of the GNF-$N$ by the value $U$ corresponding to quarter of the maximum sublattice magnetization, $S_{\rm st}=1/4$. 
We assume that in view of sharpness of the dependence $S_{\rm st}$ on $U$ near magnetic phase transition for not very small nanoflakes, the obtained values of interactions $U_c^N$ are close to the critical interaction for the magnetic transition in the limit $h\rightarrow 0$. The obtained characteristic interactions $U^{N}_{c}/t$ are presented in the Table~\ref{UcTab}. 
The obtained value of $U_c^N$ agrees well with the result of DMFT approach for $N=54$
\cite{Valli_2018}, $U_c\approx {3.6t}$ (obtained from the condition $S_{\rm st}=1/4$)   and the result of DCA approach for $N=96$,  $U_c\lesssim 3.6t$ \cite{DCA} (obtained from vanishing sublattice magnetization).


\begin{table}[]
\centering
\caption{Estimates of characteristic local interaction $U^{N}_{c}/t$, corresponding to $S_{\rm st}=0.25$.}\label{UcTab}
\resizebox{4.5cm}{!}{
\begin{tabular}{lllllll}
\hline
\hline
\multicolumn{1}{c|}{$N$} & 
\multicolumn{1}{|c}{6} & \multicolumn{1}{c}{24} & \multicolumn{1}{c}{54} & \multicolumn{1}{c}{96} &\\ \hline
\multicolumn{1}{c|}{$U^{N}_{c}/t$}  &\multicolumn{1}{c}{5.54} & \multicolumn{1}{c}{4.25} & \multicolumn{1}{c}{3.83} & \multicolumn{1}{c}{3.59}\\ 
\hline
\hline
\end{tabular}
}
\end{table}

Finally, in the inset of Fig.~\ref{GNFnLocal}, we show the linear conductance $G$ as a function of $U/t$ for the GNF-54 system. With increase of $U/t$ (and hence $S_{\rm st}$), the conductance gradually decreases. When $U \gtrsim U^{54}_{c}$, the SDW order is developed and the conductance is suppressed. The other GNF-$N$ systems under consideration produce qualitatively similar behavior of the conductance. 

\subsection{Zigzag-edge GNF-$N$ with account of non-local interaction
} 
We next study the zigzag-edge GNF-$N$ systems with both {on-site} ($U$) and non-local ($U_{i\ne j}=U^{*}_{i\ne j}/\epsilon_{\rm nl}$) electron-electron interactions.

\begin{figure}[t]
	\center{\includegraphics[width=1\linewidth]{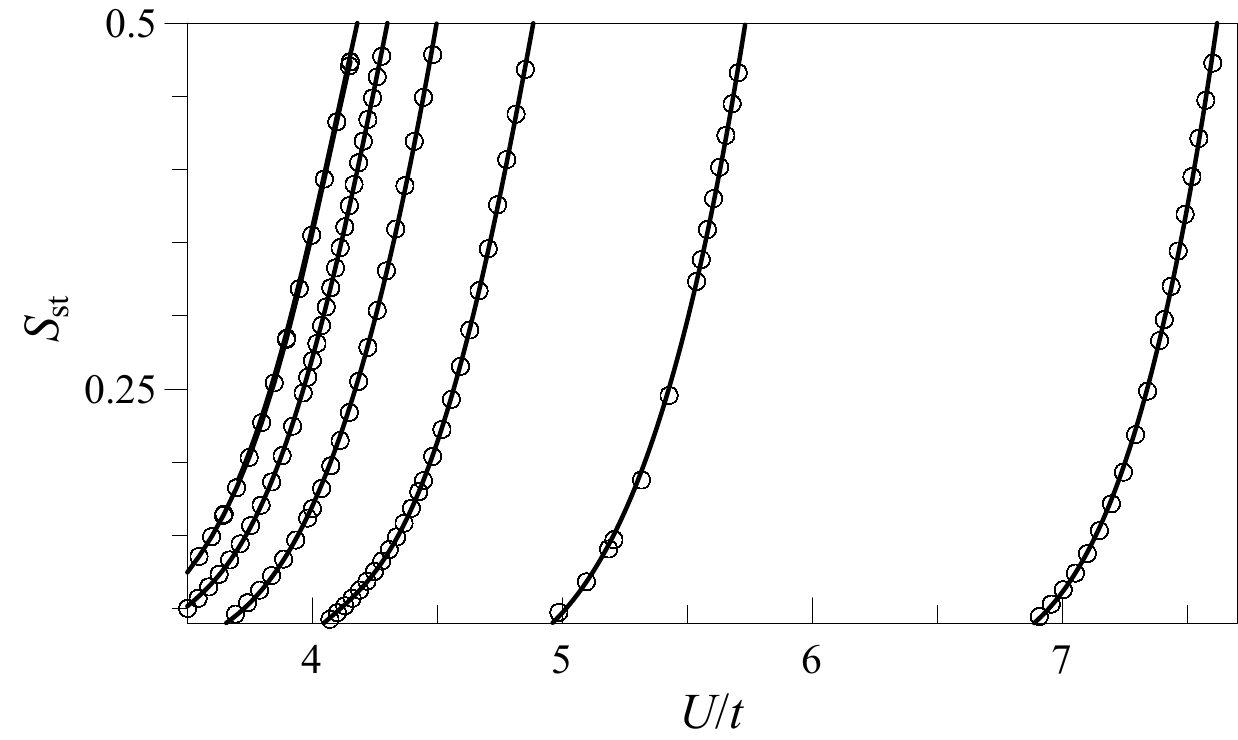}}
		\caption{The average relative staggered magnetization $S_{\rm st}$ of the GNF-54 system as a function of $U/t$ for various  $\epsilon_{\rm nl}$. From right to left: $\epsilon_{\rm nl}=0.5$, $1$, $2$, $4$, $9.8$, and $\epsilon^{-1}_{\rm nl}=0$ ($U_{i\ne j}=0$ in the latter case). The solid lines are 4th-order polynomial interpolation of the fRG data.}
	\label{SzUdiffeps54}
\end{figure}

\begin{figure}[b]
\center{\includegraphics[width=0.75\linewidth]{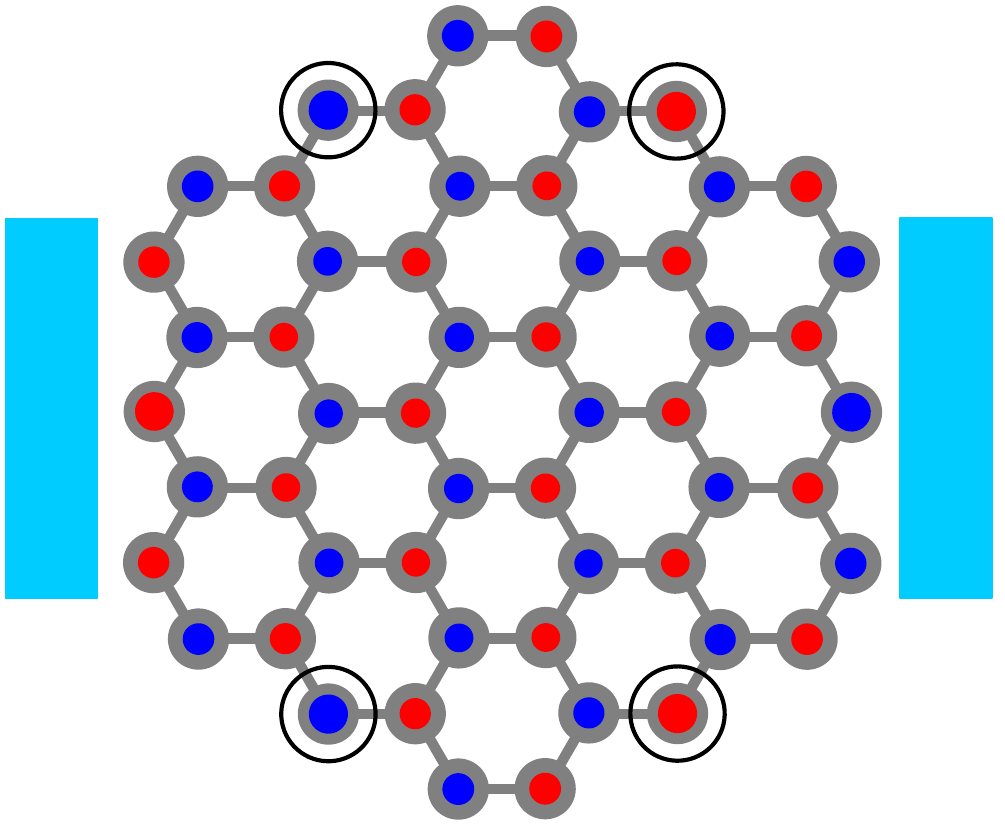}}
	\caption{(Color online) {The distribution of the magnetization $m_j=|\langle n_{j,\uparrow}-n_{j,\downarrow} \rangle|$ in GNF-54 {system} {with $U=11~{\rm eV}\approx 4.07t>U^{54}_{c}$ and $\epsilon_{\rm nl}=7$. {The {red (blue)} dots correspond to $A$ ($B$) sublattices, their size is proportional to $m_{j}$.} The {open} circles indicate the sites corresponding to the
	$\max\{m_{j}\}\approx 0.36$. The relative average staggered magnetization $S_{\rm st}\approx 0.28$. }}}
	\label{Szdistr}
\end{figure}

In Fig.~\ref{SzUdiffeps54}, we plot $S_{\rm st}$ as a function of $U/t$ for the GNF-54 system for various values of screening parameter of the non-local interaction $\epsilon_{\rm nl}$. In the case $\epsilon_{\rm nl}\gg 1$, the magnetization tends to 
that for the case when
only {on-site} interaction present. With decreasing $\epsilon_{\rm nl}$, the region with non-zero magnetization shifts to the higher values of $U/t$. The example of spin distribution {at sufficiently large $U$ in the symmetry broken phase} is shown in Fig. \ref{Szdistr}. {The spin distribution is qualitatively analogous to the one presented for the purely local interaction ($\epsilon_{\rm nl}^{-1}=0$) case in Ref.~\cite{Valli_2018}.} The other systems show similar behavior of $S_{\rm st}$ with respect to both $U/t$ and $\epsilon_{\rm nl}$.

\begin{figure}[t]
	\center{\includegraphics[width=0.9\linewidth]{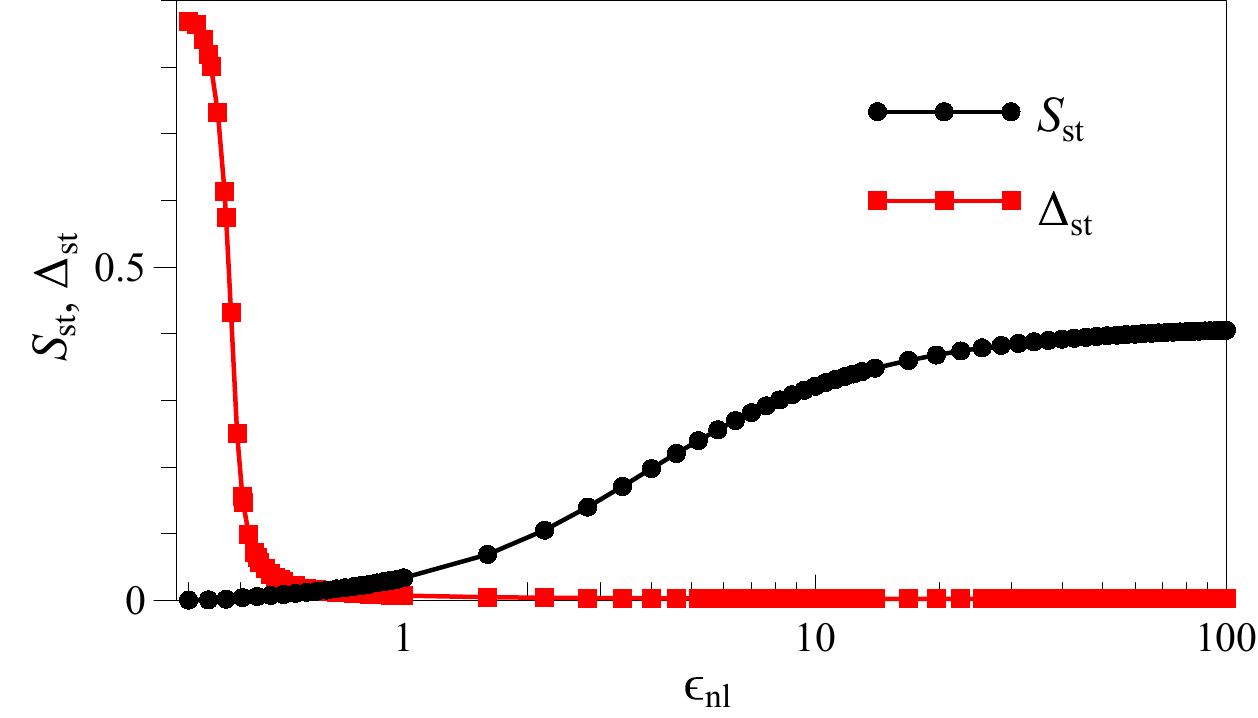}}
	\caption{(Color online) The average {relative} staggered magnetization $S_{\rm st}$ (black circles) and average {relative} difference in occupation of the sublattices $\Delta_{\rm st}$ (red squares)  as a function of $\epsilon_{\rm nl}$ (in logarithmic scale) for {GNF-54 system} with $U=11~{\rm eV}\approx4.07t$.}	\label{SzandNabP1}
\end{figure}
\begin{figure}[b]
		\center{\includegraphics[width=0.7\linewidth]{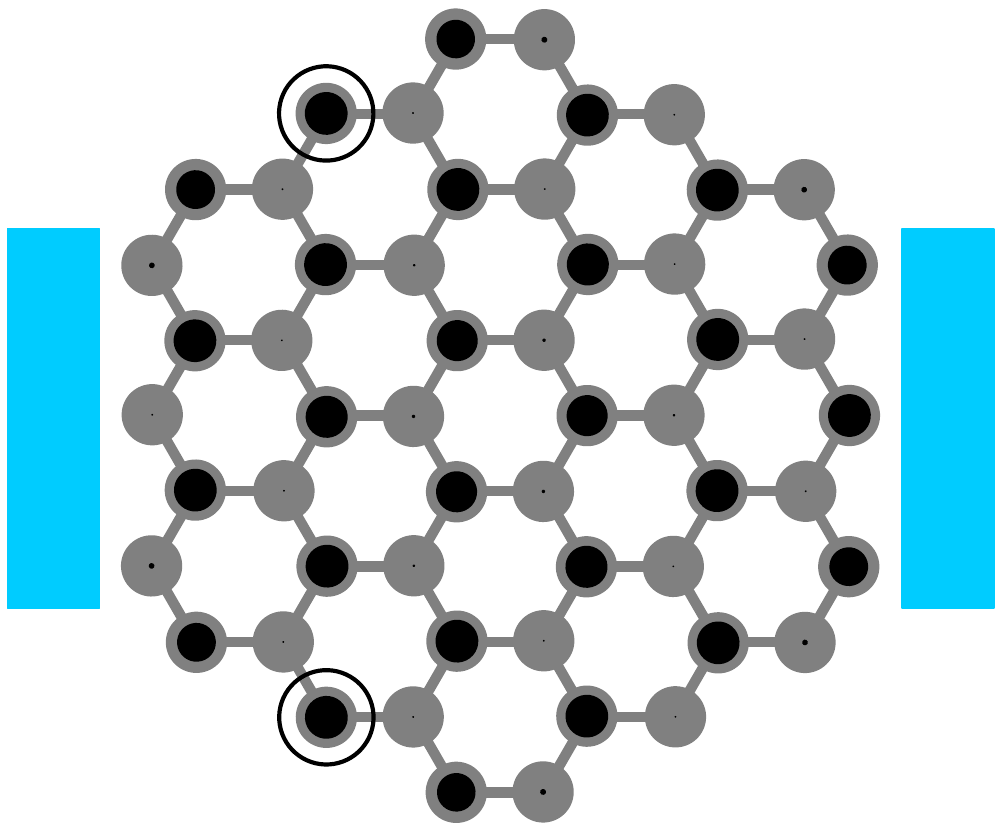}}
	\caption{(Color online) The distribution of $\langle n_{j} \rangle=\langle n_{j,\uparrow}+n_{j,\downarrow} \rangle$ for {GNF-54 system with} $U=11~{\rm eV}\approx4.07t$ and $\epsilon_{\rm nl}=0.3$. The size of the black dots is proportional to $\langle n_{j} \rangle$. The open circles indicate the sites corresponding to the $\max\{\langle n_{j} \rangle\}\approx 1.92$.}
	\label{DistNiSz2}
\end{figure}

\begin{figure}[t]
	\center{\includegraphics[width=1.0\linewidth]{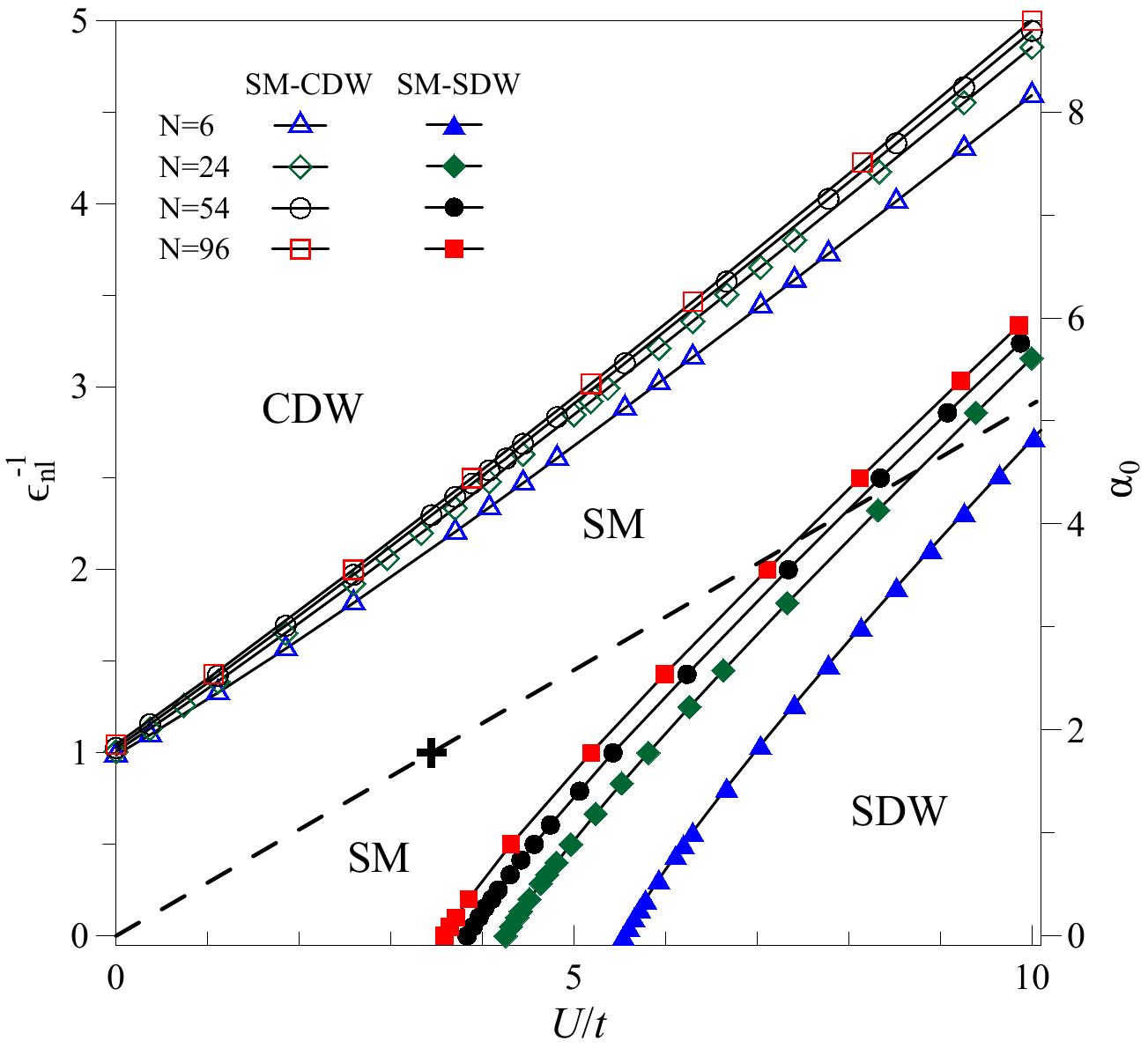}}
		\caption{(Color online) Phase diagram of the zigzag-edge GNF-$N$ system in the $\left(U/t,\epsilon_{\rm nl}^{-1} \right)$ and $\left(U/t,\alpha_0\right)$ coordinates for $N=6$ (blue triangles), $N=24$ (green diamonds), $N=54$ (black circles) and $N=96$ (red squares): SM-SDW phase transitions are denoted by solid lines with filled symbols, SM-CDW phase transitions are marked by solid lines with open symbols. The dashed line $U=U_r/\epsilon_{\rm nl}$ corresponds to simultaneos rescaling of the on-site and long-range part, considered in Ref.~\cite{Ulybyshev_2013}. The point $\epsilon_{\rm nl}=1$ and $U=9.3~{\rm eV}{\approx}3.4t$, corresponding to freely suspended graphene, is marked by the plus symbol. 
		}
	\label{PDs_main}
\end{figure}

Fig.~\ref{SzandNabP1} shows the average {relative} difference between the occupation of the sublattices $\Delta_{\rm st}$ and the average {relative} staggered magnetization $S_{\rm st}$ as a function of $\epsilon_{\rm nl}$, for GNF-54 system with $U=11~{\rm eV}\approx 4.07t$. When the magnetization $S_{\rm st}$ is almost suppressed, $\Delta_{\rm st}$ increases monotonously with decreasing $\epsilon_{\rm nl}$. In the limit $\epsilon_{\rm nl} \ll 1$ we have $\Delta_{\rm st}\approx 1$ and $S_{\rm st}\approx 0$, which corresponds to the CDW order of the system. The establishing of the CDW order for $\epsilon_{\rm nl}\lesssim 1$ is clearly seen from the checkerboard distribution of $\langle n_{j} \rangle=\langle n_{j,\uparrow}+n_{j,\downarrow} \rangle$ (see, e.g., Fig.~\ref{DistNiSz2}). In particular, for $\epsilon_{\rm nl}=0.3$ we have $\langle n_{j\in B} \rangle\approx 2$ and $\langle n_{j\in A} \rangle\approx 0$ for GNF-54 system.
Thus, the decrease of $\epsilon_{\rm nl}$, which corresponds to the increase of the long-range interaction, drives the phase transition from the SM state ($S_{\rm st}\approx 0$, $\Delta_{\rm st}\approx 0$) to the CDW one ($S_{\rm st}\approx 0$, $\Delta_{\rm st}\approx 1$). We find that the SM-CDW phase transition takes place for all GNF-$N$ systems and an arbitrary value of on-site interaction $U$ parameter (including $U=0$ case), while SM-SDW one occurs for $U>U^{N}_{c}$.

To obtain the SM-SDW phase boundaries, we again define the {characteristic} $U/t$ for fixed $\epsilon_{\rm nl}$ as the value corresponding to $S_{\rm st}=1/4$.
The resulting  dependence of inverse critical screening $\epsilon_{\rm nl}^{-1}$ and corresponding graphene's ``fine structure" constant $\alpha_{0}=e^2/(\epsilon_{\rm nl}\epsilon_{\rm eff} v^0_F)$,   
where $v_F^0=3at/2$ is the bare Fermi velocity, on the {obtained values} of $U/t$
are shown in Fig.~\ref{PDs_main}. For $\epsilon_{\rm nl}\gg 1$ the long-range part of the non-local potential is small and the on-site interaction $U$ plays the main role. In this limit, the on-site interaction $U\gtrsim U^{N}_{c}$ (as for the $\epsilon_{\rm nl}^{-1}=0$ case) leads to the SDW order of the systems. When $\epsilon_{\rm nl}$ is sufficiently small, the SDW state becomes unstable and the SM state occurs for all GNF-$N$ systems. 
With increase of the the size of the system $N$ the {critical value of the parameter $\epsilon_{\rm nl}$ decreases}.

Analogously to the case of SM-SDW phase transition, we define the SM-CDW phase boundary as a line  $\epsilon_{\rm nl}^{-1}( U/t)$ at which $\Delta_{\rm st}=1/4$. 
Our results for the SM-SDW and SM-CDW phase-transition lines for different GNF-$N$ systems are also summarized in Fig.~\ref{PDs_main}. For $U<U^{N}_{c}$, the GNF-$N$ system undergoes the phase transition from the SM to the CDW ground state induced by changing $\epsilon_{\rm nl}$ parameter. When $U>U^{N}_{c}$, the SDW phase occurs and with increasing $\epsilon_{\rm nl}$ from value $\epsilon_{\rm nl}=0$ the system undergoes two phase transitions: first, from the CDW to the SM and, second, from the SM to the SDW phase. 
The calculated SM-CDW phase transition lines $\epsilon_{\rm nl}^{-1}(U/t)$  are very close to each other (except $N=6$ case) and are well approximated by linear dependencies.

The important feature of the phase diagram of Fig.~\ref{PDs_main} is strong increase of critical  {$\alpha_0$} of charge instability with increase of $U$. As a result, in contrast to the case of onsite and nearest neighbor interaction \cite{Herbut_2006, DCA} there is a wide region of the phase diagram with no instability.
The GNF-$N$ system with the realistic non-local interaction (for $U=U_r$ and $\epsilon_{\rm nl}=1$) falls into this region and therefore corresponds to the SM ground state of the system (marked with the {plus symbol} in Fig.~\ref{PDs_main}). Moreover, as can be seen from Fig.~\ref{PDs_main}, all considered GNF-$N$ systems with $\epsilon_{\rm nl}>1$ are far from both SM-SDW and SM-CDW phase-transition lines. 

\begin{figure}[t]
	\center{\includegraphics[width=1.0\linewidth]{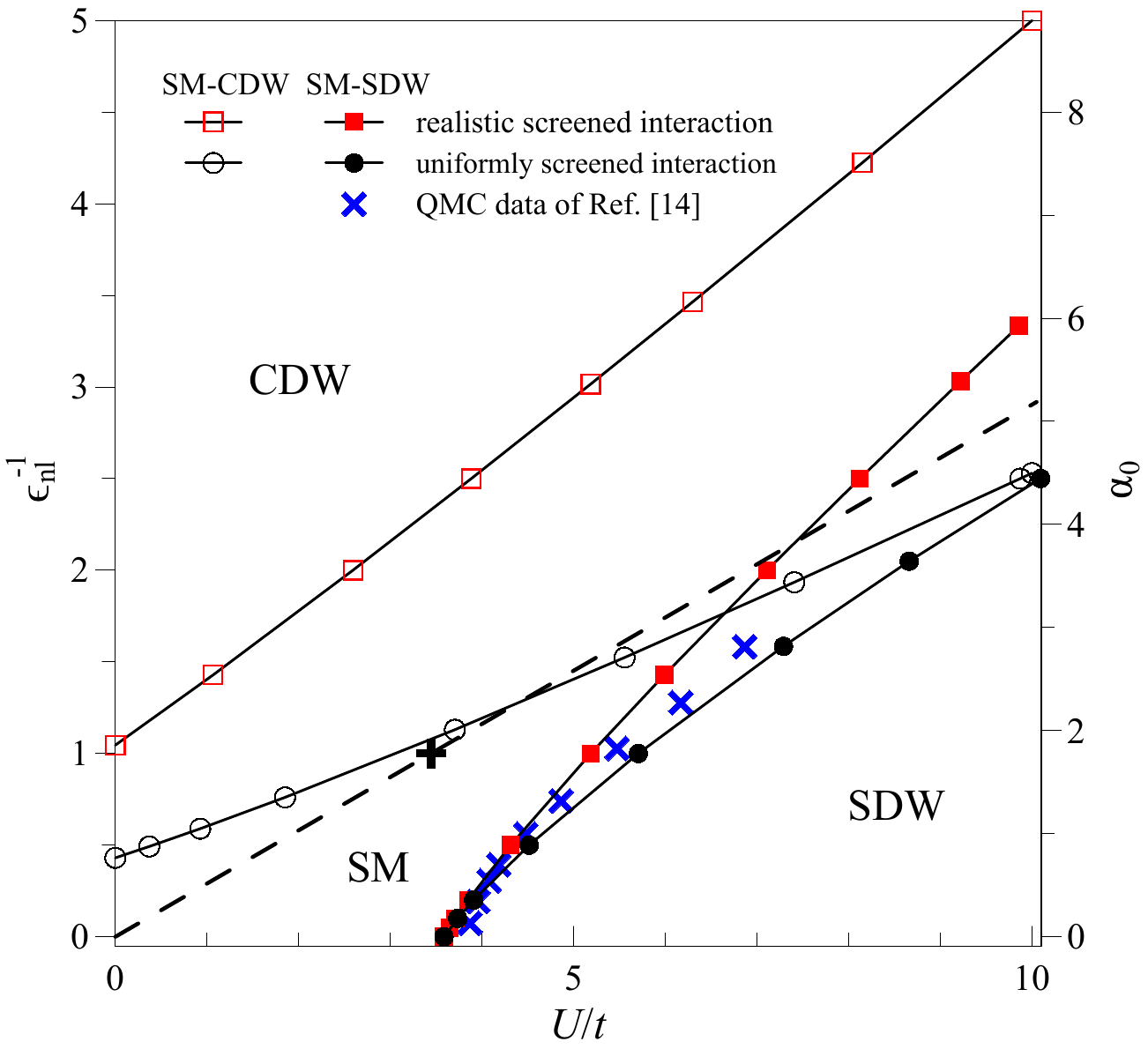}}
			\caption{(Color online) Critical $\epsilon_{\rm nl}^{-1}$ and corresponding fine structure constant ${\alpha_{0}=e^2/(\epsilon_{\rm nl}\epsilon_{\rm eff}  v^0_F)}$ for SM-SDW (solid lines with filled symbols) and SM-CDW (solid lines with open symbols) phase transitions as functions of $U/t$ for GNF-$96$ system with the realistic screened $U_{i\neq j}= U^*_{ij}/\epsilon_{\rm nl}$ (red squares) and uniformly screened $U_{i\neq j}=e^2/(\epsilon_{\rm nl}\epsilon_{\rm eff}r_{ij})$ (black circles) form of non-local interaction. The blue crosses are data obtained from scaling analysis of QMC results in Ref.~\cite{Tang_2018}. The dashed line $U=U_r/\epsilon_{\rm nl}$ corresponds to simultaneous screening of local and non-local interaction. The point $\epsilon_{\rm nl}=1$ and $U=9.3~{\rm eV}{\approx}3.4t$, corresponding to freely suspended graphene, is marked by the plus symbol.
			}
			\label{PDs_AFM_PM}
\end{figure}

To provide an insight into importance of screening of nearest- and next-nearest neighbor Coulomb interactions by $\sigma$ bands, which yields the difference of the considered non-local potential $U^*_{ij}$ of Refs. \cite{Wehling_2011,Ulybyshev_2013} from the bare Coulomb interaction, 
in Fig.~\ref{PDs_AFM_PM} we compare the above discussed fRG results 
to the results for the uniformly screened Coulomb interaction $U_{i\neq j}= e^2/(\epsilon_{\rm nl}\epsilon_{\rm eff}r_{ij})$. 
One can see that realistic screening of Coulomb interaction, having smaller nearest- and next-nearest neighbor interaction, only moderately increases critical constant $\alpha_0$ for SDW instability, but strongly enhances critical {non-local} interaction for the charge instability. Without this enhancement freely suspended graphene nanoflakes (as well as an infinite graphene sheet) would be on the verge of the charge instability. Although this effect was qualitatively discussed previously in Ref. \cite{Ulybyshev_2013}, the position of only spin, and not charge instability, with account of screening effects was analyzed in that study. The obtained critical non-local interaction of charge density wave at $U=0$ for GNF-96 system corresponds to {$\epsilon_{\rm nl}\approx 0.96$}, i.e. {$\alpha_0^c\approx 1.86$} for realistic non-local interaction and {$\epsilon_{\rm nl}\approx 2.32$}, i.e. {$\alpha_0^c \approx 0.77$} for uniformly screened interaction. The latter value 
is not far from the critical $\alpha_0\simeq (0.9-1.1)$ for charge instability in an infinite plane, obtained by QMC in Ref. \cite{Drut} and mean-field approximation with dynamic renormalized Coulomb interaction \cite{Khveschenko_d,Gusynin_d}. 

\begin{figure}[t]
	\center{\includegraphics[width=0.9\linewidth]{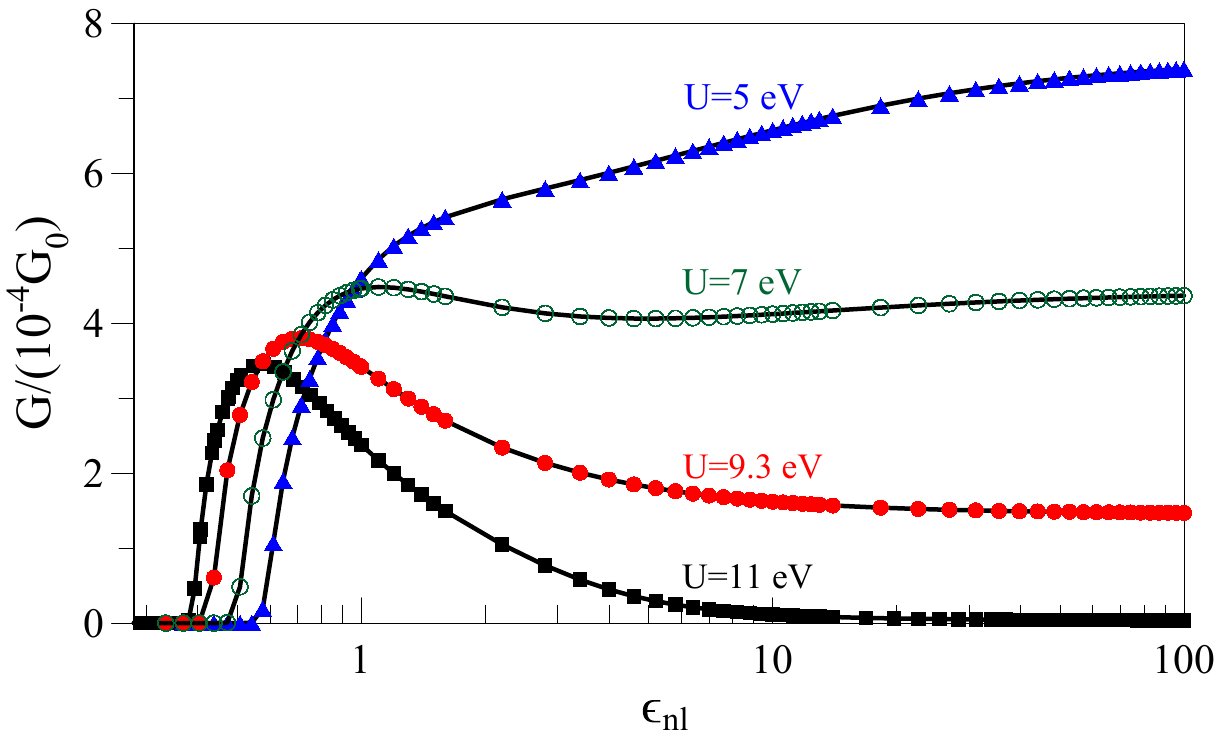}}
	\caption{(Color online) The  linear conductance $G$ as a function of $\epsilon_{\rm nl}$ {(in logarithmic scale)} for {GNF-54 system with} $U=5~{\rm eV}\approx 1.85t$ (blue triangles), $U=7~{\rm eV}\approx {2.6t}$ (dark green {open} circles), $U=9.3~{\rm eV}\approx 3.44t$ (red {filled} circles), $U=11~{\rm eV}\approx 4.07t$ (black squares).}
	\label{G_dU_v1}
\end{figure}

For realistic non-local interaction the SDW instability of GNF is reached along the line $U=U_r/\epsilon_{\rm nl}$, corresponding to simultaneous rescaling of the on-site and non-local interaction, {$U_{ij}=U^{*}_{ij}/\epsilon$, the parameter $\epsilon=\epsilon_{\rm nl}$ can be viewed as the dielectric permittivity of the medium surrounding GNF.} The crossing of the boundary of SDW instability with this line corresponds for GNF-96 to the physically unreachable value {$\epsilon={0.46}$}, {which is very} close to that for infinite sheet, obtained in Ref. \cite{Ulybyshev_2013}, {see Ref. \cite{Note2}}. At the same time, for uniformly screened interaction the SDW instability is not obtained at all along the same path. The result for the critical interaction constant $\alpha_0(U/t)$ for GNF-$96$ system, corresponding to SDW instability for uniformly screened Coulomb interaction, is also compared in Fig. \ref{PDs_AFM_PM} to that from QMC analysis in large systems ~\cite{Tang_2018,Note}. The obtained fRG result provides quantitative agreement with the scaling analysis of QMC data.


\begin{figure}[t]
	\center{\includegraphics[width=0.95\linewidth]{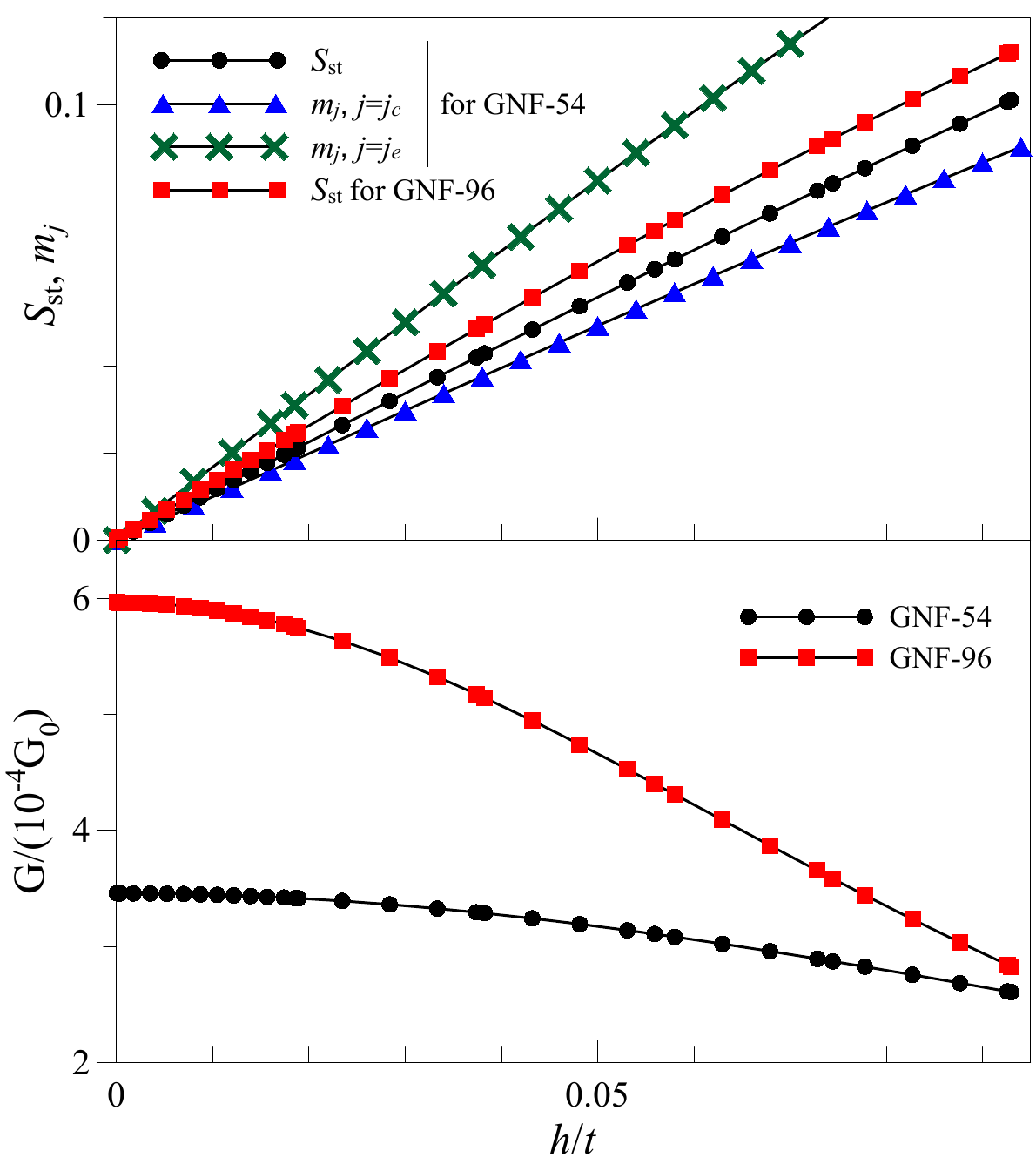}}
	\caption{(Color online) Upper part: The average relative staggered magnetization $S_{\rm st}$ for GNF-$54$  (black circles) and GNF-$96$  (red squares), together with the magnetization of GNF-$54$ system $m_j={|}\langle n_{j,\uparrow}-n_{j,\downarrow}\rangle{|}$ at the center site $j_c$ (blue triangles) and at the center of edge $j_e$ (green crosses) {as a function of $h/t$}. Lower part: {The} {linear} conductance $G$ as a function of $h/t$ for {the GNF-$54$ (black circles) and GNF-$96$ (red squares) systems. The realistic non-local potential {($U=U_r$, $\epsilon_{\rm nl}=1$)} is considered.} 
	}
	\label{SzHreal}
\end{figure}

In Fig.~\ref{G_dU_v1} the linear conductance $G$ of the GNF-54 system at $T=0$ is plotted as a function of $\epsilon_{\rm nl}$. For strong non-local interaction ($\epsilon_{\rm nl}\ll 1$) the conductance is suppressed due to formation of CDW. {For $U>U_{c}^{54}\approx 3.8t$ (see for e.g. the plot $G(\epsilon_{\rm nl})$ for $U=11~{\rm eV}\approx 4.07t$)} the conductance is suppressed also for weak $(\epsilon_{\rm nl}\gg 1)$ non-local interaction due to the development of the SDW order. In the latter case at some intermediate $\epsilon_{\rm nl}$ at which $S_{\rm st}, \Delta_{\rm st}\approx 0$  (see Fig.~\ref{SzandNabP1}) the conductance has a maximum. 
When $U\lesssim U^{54}_{c}$ the SDW phase does not occur and peak in the conductance gradually disappears with decreasing $U$, such that $G$ becomes a monotonous function of $\epsilon_{\rm nl}$. In the latter case, opposite to the $U>U^{54}_{c}$ case, the conductance tends to a nonzero value in the limit $\epsilon_{\rm nl}\gg 1$. 
We have found that the above behavior of the conductance is generic for all GNF-$N$ systems under consideration.

{Finally, in Fig.~\ref{SzHreal} we show the magnetic field dependence of the average relative staggered magnetization $S_{\rm st}$ and linear conductance $G$ of the GNF-$54$ and GNF-$96$ systems with the realistic non-local potential ($U=U_r$, $\epsilon_{\rm nl}=1$). For both systems and small magnetic fields $h\lesssim {0.02t}$ the sublattice magnetization is well fitted by the linear function $S_{\rm st}\left(h\right)=\chi^{N}h$, 
which confirms the SM nature of the GNFs for the realistic parameters. The parameter $\chi^{N}$ can be considered as the paramagnetic susceptibility.} {We find $\chi^{54}\approx 1.13/t$ and $\chi^{96}\approx 1.32/t$ for the GNF-54 and GNF-96 system, respectively, which corresponds to a paramagnetic state, albeit with pronounced spin correlations.} {To characterize the distribution of spin order in finite magnetic field, we also present the results for the absolute value of the magnetization $m_j=|\langle n_{j,\uparrow}-n_{j,\downarrow}\rangle|$ at the center site $j_c$  and at the center of edge $j_e$ of GNF-54 system. One can see that the magnetization at the edge is substantially larger than that in the center, and also substantially different from the staggered relative magnetization $S_{\rm st}$, due to formation of the edge states.  }    {The linear conductance $G$ depends monotonously on the magnetic field and becomes almost constant for relatively small magnetic fields $(h/t\lesssim 0.02)$.}

\begin{figure}[b]
       \center{\includegraphics[width=0.8\linewidth]{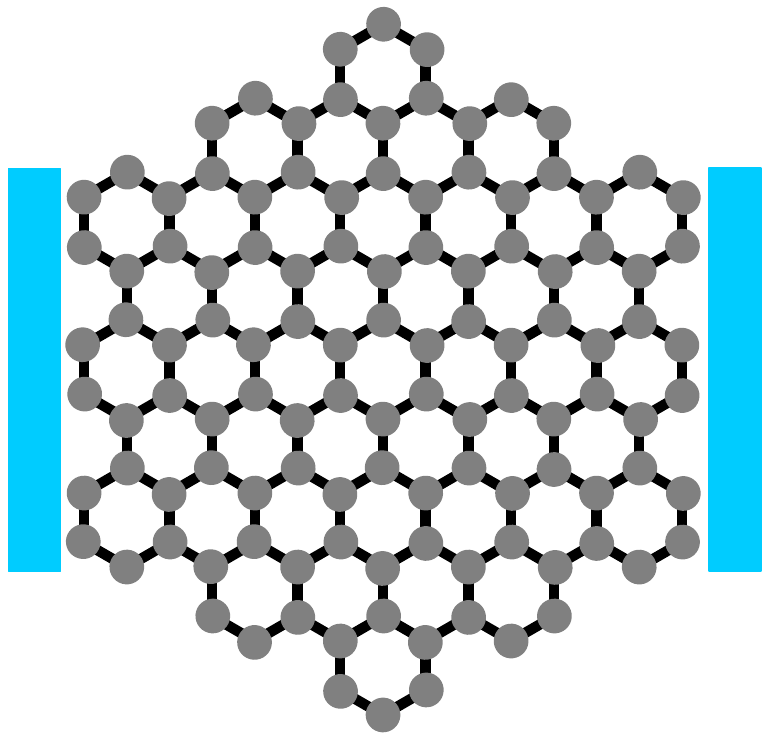}}
		\caption{{(Color online)} Armchair-edge GNF system with $N=114$. 
		The left and right leads are shown schematically by rectangles.}
\label{ArmchairSystemsPic}
\end{figure}

\subsection{Comparison to armchair GNFs}
\begin{figure}[t]
	\center{\includegraphics[width=1.0\linewidth]{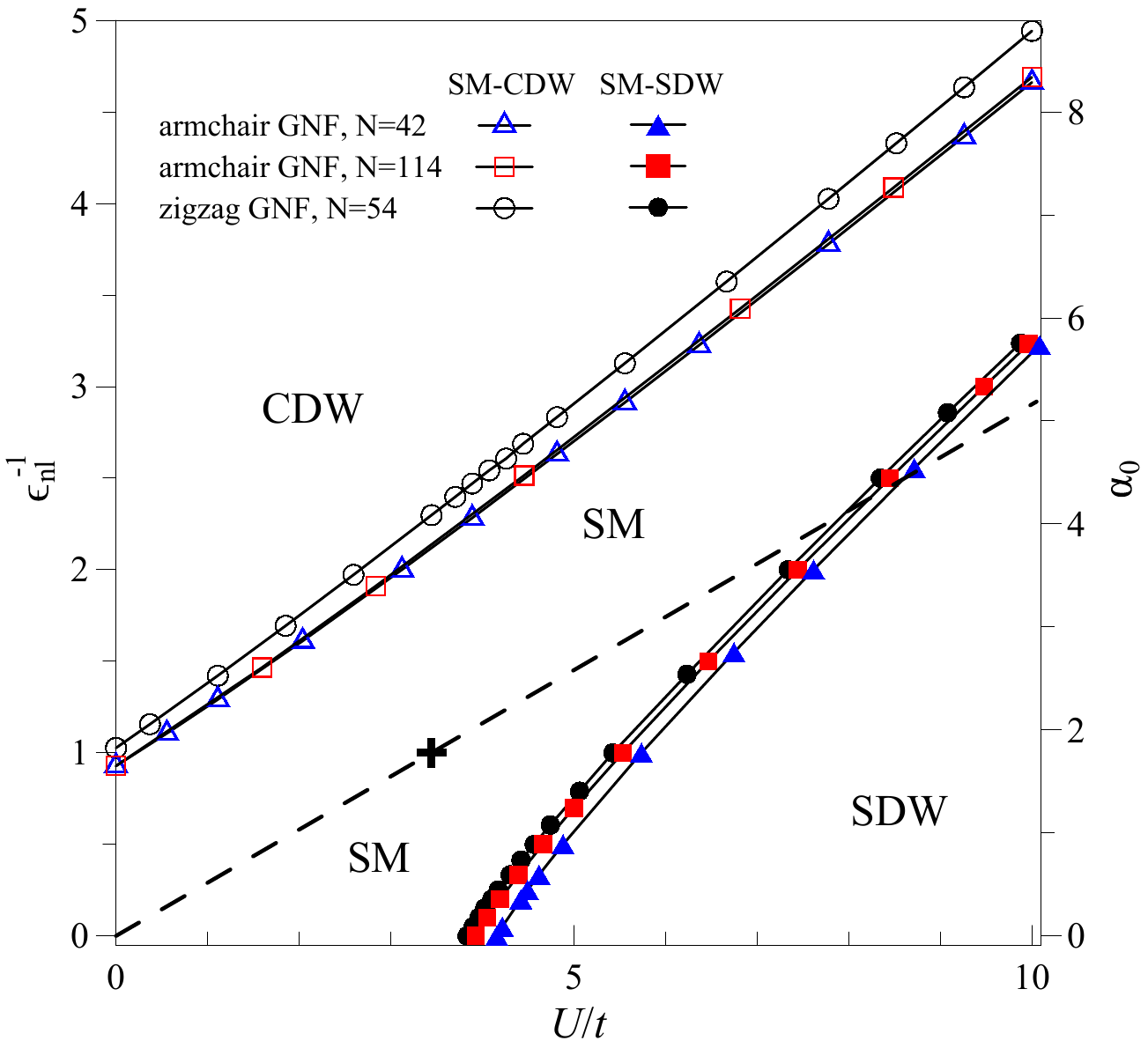}}
		\caption{(Color online) Phase diagrams  for armchair-edge GNF-42 (blue triangles), armchair-edge GNF-114 (red squares), compared to the zigzag-edge GNF system with $N=54$ (black circles): SM-SDW phase transitions are denoted by solid lines with filled symbols, SM-CDW phase transitions are marked by solid lines with open symbols. The dashed line $U=U_r/\epsilon_{\rm nl}$ corresponds to simultaneos rescaling of the on-site and long-range part, considered in Ref.~\cite{Ulybyshev_2013}. The point $\epsilon_{\rm nl}=1$ and $U=9.3~{\rm eV}{\approx}3.4t$, corresponding to freely suspended graphene, is marked by the plus symbol. 
		}
	\label{armchairPDs}
\end{figure}
{
In the following we compare the obtained results to those for the armchair-edge GNFs with $N=42$ and $N=114$ atoms (see Fig.~\ref{ArmchairSystemsPic}). To define the SM-CDW and SM-SDW phase-transition lines for armchair-edge GNFs we use the same procedure as in Section~IIIB. The corresponding lines of phase transitions for these systems are presented and compared to zigzag-edge GNF-54 in Fig.~\ref{armchairPDs}. The position of the SM, CDW, and SDW phases in the $\left(U/t,\epsilon_{\rm nl}^{-1} \right)$ coordinates remains qualitatively the same as in the case of zigzag-edge GNFs. Furthermore, as for the zigzag-edge GNFs with $N>6$ (see Fig.~\ref{PDs_main}) the SM-SDW (SM-CDW) phase-transition lines for GNFs with armchair edges are (very) close to each other even if $N$ are substantially different. However, for fixed $U$ both SDW and CDW phase transitions for armchair GNFs take place at somewhat smaller values of non-local interaction (higher $\epsilon_{\rm nl}$) than the corresponding transitions in zigzag-edge GNFs with close $N$. 
For example, the SM-SDW phase transition line for the armchair-edge $N=114$ system is quantitatively closer to the one for the zigzag-edge GNF-54 rather than to the one obtained for the zigzag-edge GNF-96 system.} 

{The conductance for the armchair-edge geometry (not shown) is reduced by about an order of magnitude compared to the conductance of zigzag-edge systems with close size. This stresses importance of interference effects for GNF conductance. However, the general relations between the conductance and magnetic states of the GNF systems revealed in the present study are preserved also for an armchair edge geometry. 
}
\section{Conclusion}
In this paper, we have investigated magnetic, charge, and transport properties of hexagonal graphene nanoflakes (GNFs) connected to two metallic leads. Both on-site $U$ and long-range interaction effects in GNFs are taken into account. Using the functional renormalization group method we have calculated the average relative staggered magnetization, the average relative difference between the occupation of sublattices, and the linear conductance. The ground-state phase diagrams at half-filling are obtained for the GNF systems with realistic screened, as well as the uniformly screened long-range Coulomb interaction. The obtained phase diagram in $(U/t,\epsilon_{\rm nl}^{-1})$ coordinates, where parameter $\epsilon_{\rm nl}$ rescales the strength of the non-local interaction, is shared by three phases: the semimetal (SM), the spin-density-wave (SDW), and the charge-density-wave (CDW) phase. 

At first, we have analyzed the zigzag-edge GNFs with screened realistic  Coulomb interaction of Refs.~\cite{Ulybyshev_2013,Wehling_2011}. We showed that with increasing size of the GNF, the phase boundary between the SM and SDW phases shifts to higher (lower) values of critical long-range (on-site) interaction strength. The transition line between the SM and CDW phases is almost linear in $(U/t,\epsilon_{\rm nl}^{-1})$ coordinates and weakly depends on the GNF size. We have found that for the realistic long-range interaction parameters freely suspended GNFs are far from both SM-SDW and SM-CDW phase-transition boundaries and belong to the SM phase. The estimated critical values of the on-site interaction $U_{c}^{N}$, corresponding to the SDW instability for purely local interaction, agree well with the result of the DMFT approach for $N = 54$~\cite{Valli_2018} and the result of the DCA approach for $N = 96$~\cite{DCA}. 

Then, for comparison, we have presented the ground-state phase diagram for the zigzag-edge GNF system with $N=96$ and the uniformly screened Coulomb potential. For the SM-SDW phase transition, we have found moderate suppression of the critical long-range interaction strength in comparison to the one obtained for the realistically screened Coulomb interactions. In contrast, the transition line between the SM and CDW phases is rather different for the realistic and uniformly screened long-range Coulomb interactions. The realistic screening of Coulomb interaction by $\sigma$ bands causes a strong enhancement of the critical value of long-range interaction needed to stabilize the CDW state. This results in a substantially wider region of stability of the SM phase for the case of the realistic non-local potential. In particular, the critical non-local interaction of the CDW phase at $U=0$ is sufficiently larger for the realistic potential than the corresponding value for the uniformly screened potential. The latter value is consistent with the results of QMC in Ref. \cite{Drut} and mean-field approximation with dynamic renormalized Coulomb interaction \cite{Khveschenko_d,Gusynin_d}.  The result on the SM-SDW transition line agrees well with the scaling analysis of QMC data~\cite{Tang_2018}.

Finally, we have shown that the behavior of the linear conductance $G$ of the GNF system has a close connection with its magnetic or charge order. In particular, the linear conductance is strongly suppressed both in the SDW and CDW phases. The linear conductance as a function of $\epsilon_{\rm nl}$ exhibits a peak for $U>U^{N}_{c}$, corresponding to the SM phase of the system. At the same time for $U<U^{N}_{c}$, when the SDW phase does not occur,  the peak is absent and the conductance $G(\epsilon_{\rm nl})$ shows a monotonous behavior. 

{Analysis of  GNFs with armchair edges shows that for fixed local interaction $U$ both SDW and CDW phase transitions for these systems take place at somewhat smaller values of non-local interaction (higher $\epsilon_{\rm nl}$) than the corresponding transitions in zigzag-edge GNFs with close $N$.}

We emphasize that the present fRG study is limited to the case of an ideal (non-distorted) graphene lattice, and the Kekul{\'e} bond order phases~\cite{Gutierrez_2016, Xu_2018} do not appear in the phase diagrams of GNFs. The possibility of these phases in GNFs can be also considered within the fRG approach when an initial small distortion of the hopping matrix elements between A and B sublattices, playing the role of symmetry breaking perturbation, is introduced. {Another important issue is the consideration of disorder effects, which may play a significant role in GNFs, as is the case for graphene~\cite{Ostrovsky_2006, Ostrovsky_2007, Ostrovsky_2010, Katanin_2013, Sbierski_2017}. The effects of disorder are expected to reduce the tendency towards charge and spin ordering in GNFs.}
{A detailed analysis of these issues is beyond the scope of the present investigation but would be an interesting subject for future studies.}\par

We also note that the results of the present paper are obtained at half-filling and for an equilibrium state of the GNF systems. However, the functional renormalization group method used in our study can be straightforwardly applied beyond both these limitations. In this perspective, investigation of magnetic and charge properties of GNF for finite gate and bias voltages has to be performed. Apart form that, study of other carbon nanoobjects, e.g. carbon nanotubes, is is of certain interest.

\section*{Acknowledgements} The authors are grateful to A. Valli and M. Capone for stimulating discussions. The work was performed within the state assignment from the Ministry of Science and Higher Education of Russia (theme "Quant" AAAA-A18-118020190095-4) and partly supported by RFBR grant 20-02-00252a. A. A. Katanin also  acknowledges the financial support from the Ministry of Science and Higher Education of the Russian Federation (Agreement No. 075-15-2021-606). The calculations were performed on the Uran supercomputer at the IMM UB RAS.

\end{document}